*Article*

# Sustainable Banking; Evaluation of the European Business Models

Saeed Nosratabadi [1], Gergo Pinter [2], Amir Mosavi [3,4,5*] and Sandor Semperger [6]

[1] Faculty of Humanities and Social Sciences, Oxford Brookes University, Oxford OX30BP, UK; saeed.nosratabadi@phd.uni-szie.hu

[2] John von Neumann Faculty of Informatics, Obuda University, 1034 Budapest, Hungary; pinter.gergo@nik.uni-obuda.hu

[3] Institute of Research and Development, Duy Tan University, Da Nang 550000, Vietnam

[4] Thuringian Institute of Sustainability and Climate Protection, 07743 Jena, Geramny

[5] Faculty of Health, Queensland University of Technology, 130 Victoria Park Road, Queensland 4059, Australia

[6] Institute of Automation, Kalman Kando Faculty of Electrical Engineering, Obuda University, 1034 Budapest, Hungary; semperger.sandor@kvk.uni-obuda.hu

\* Correspondence: amirhoseinmosavi@duytan.edu.vn

**Abstract:** Sustainability has become one of the challenges of today's banks. Since sustainable business models are responsible for the environment and society along with generating economic benefits, they are an attractive approach to sustainability. Sustainable business models also offer banks competitive advantages such as increasing brand reputation and cost reduction. However, no framework is presented to evaluate the sustainability of banking business models. To bridge this theoretical gap, the current study using A Delphi-Analytic Hierarchy Process method, firstly, developed a sustainable business model to evaluate the sustainability of the business model of banks. In the second step, the sustainability performance of sixteen banks from eight European countries including Norway, The UK, Poland, Hungary, Germany, France, Spain, and Italy, assessed. The proposed business model components of this study were ranked in terms of their impact on achieving sustainability goals. Consequently, the proposed model components of this study, based on their impact on sustainability, are respectively value proposition, core competencies, financial aspects, business processes, target customers, resources, technology, customer interface, and partner network. The results of the comparison of the banks studied by each country disclosed that the sustainability of the Norwegian and German banks' business models is higher than in other counties. The studied banks of Hungary and Spain came in second, the banks of the UK, Poland, and France ranked third, and finally, the Italian banks ranked fourth in the sustainability of their business models.

## 1. Introduction

Sustainable business models (SBMs) are tools allowing the businesses to meet the environmental, the social, and the economic benefits simultaneously [1]. Although the nature of the banking sector, as the backbone of the finance industry, requires fewer natural resources, carbon dioxide emissions, and waste management activities in comparison with the manufacturing companies, there is much evidence in the literature that sustainability and going green will lead the service sectors toward profitability [2,3]. Leveraging of the business reputation [4,5] brand differentiation [6], cost reduction through the management of energy and water consumption [7] increasing employee satisfaction and retention [8,9], directing the industry toward best practices [10], responding the needs of customers with eco-preferences [11] are the benefits named in the literature of service sectors to apply sustainability principles. As Yip and Bocken [12] articulate that the banks, as an intermediary, facilitate the sustainable development processes.

There are many strategies proposed for service sectors to implement sustainability principles. Waste management (e.g., [13,14]), water (e.g., [15]) and energy consumption management (e.g., [16–18]), and powerful employees (e.g., [19,20]) are strategies applied in service sectors to deal with sustainability concerns. To obtain the competitive advantage of sustainability, complex processes are needed to manage the transition to sustainability. Such a transition can be achieved through innovative services, innovation in service creation and service delivery approaches, and new forms of business partnerships [21]. Going sustainable slakes the cost to nature and increases sustainable economy capacities [22]. Accordingly, strategies that are able to promote the sustainable value creation [23] and the cost leadership [24] are widely recommended for service sectors to transition toward sustainability. The SBM arms the managers to design sustainable values and provides a big picture of the business operations. According to Shen, Shuai, Jiao, Tan, and Song [25], banks always encounter with financial, social, and environmental risks while only financial risks are evaluated and calculated. Jeucken [26] believes that sustainable banks provide their customer's sustainable products that encourage socially responsible investment. In addition, the stakeholders judge the banks according to their social performance, along with their economic performance. Therefore, capital markets give awards to policies that are socially and environmentally responsible. Thence, such demands from the clients and the other stakeholders stimulate the firms to focus more on sustainable services through redefining their business logic and integrating sustainability in their policies [27]. Subsequently, the banks establish sustainable development through their strategies [28]. The current literature focuses on strategies such as socially responsible investment (i.e., [29]), corporate social responsibility (i.e., [30]), going green (i.e., [31]) to meet the sustainable development goals. However, far too little attention has been paid to the importance of the role of the business models in achieving the sustainability goals in the banking sector whilst many researchers consider the business models as tools for implementing the strategies (e.g., [32,33]), therefore, designing an SBM provides the opportunities for the banks to execute their sustainability strategies. An SBM is a tool that incorporates multi-stakeholder benefits and innovation to create and deliver sustainable values to implement the principles of sustainability and make money simultaneously [34]. The phrase of multi-stakeholder is similar to the triple bottom line where the stakeholders are the environment, society and business owners (shareholders) [34]. Therefore, Nosratabadi et al. [1] argue that an SBM is designed in a way to meet the benefits of the environment, society and the business (economic benefit of a business) simultaneously. The SBMs provide solutions to the businesses to make money through decreasing the harmful effects of economic activities on the environment and the society. Besides, Anwar [35] considers sustainable business models as sources of competitive advantage. Accordingly, it can be interpreted that an SBM constitutes a set of elements and interrelations among them to create and deliver sustainable values for multi-stakeholders.

The European countries are leading in sustainability and they use very sophisticated financial systems. Moving toward sustainability has been very challenging for European banks. Therefore, the main objective of the current study is to assess the sustainability performance of banks in Europe. Sustainable Banking Business Model (SBBM) sees the achievement of sustainability goals in the business model. In the design of the SBBM, the achievement of sustainability goals must also be considered. However, no research has been found that develops a theoretical framework to design an SBBM. To bridge this gap in the literature, the current research takes one pace backward to develop a theoretical framework for designing an SBBM. In other words, this study aims to find the elements which support the organizational decision-makers to design an SBBM. Besides, identifying the effectiveness and contribution of each business model element in meeting the sustainability goals provide insights for the decision-makers to optimize their performance in meeting economic, environmental and social benefits. Since it illustrates how changes in one element of a business model affect their sustainability performance. Thus, the first sub-objective of this study is to find the elements of an SBM for the banking sector. The second sub-objective of the study is to distinguish the importance of each element in sustainability. The third sub-objective of the study is to evaluate and compare the sustainable performance of the business model of banks in different European countries. To do so, a wide literature is reviewed to find the elements of an SBBM. Then, a Delphi method is conducted to identify the proper elements for the banking sector. In the next phase of the study, the identified elements are prioritized by the experts, and the results are analyzed by the analytic hierarchy process method based on their potential in reaching the sustainability goals.

## 2. Research Background

Evaluation of financial performance versus social and environmental performance has turned into a hot topic during the last few years. There are many pieces of evidence proving the positive impact of corporate environmental performance on corporate financial performance (e.g., [36,37]). Besides, Margolis et al. [38] elaborate that corporate social performance affects positively the corporate financial performance. Orlitzky et al. [39] and Friede et al. [40], based on the systematic literature review, provide evidence debating social responsibility and environmental responsibility increase financial performance. Despite benefits such as leveraging the business reputation [5,6] brand differentiation [7], cost reduction thrills the banks to move toward sustainability, Wu and Shen [41] believe that improving the financial performance is the major driving force of the banks to become sustainable. Following the current solutions in the literature for the transition of banking toward sustainability are provided.

*2.1. Banking and Sustainability*

A glance at the literature reveals research projects that consider sustainable banks as banks that have been made responsible for the environment and society at the core of their business and mission [42,43]. The evolution of sustainable banking began when banks held themselves accountable to society and concepts such as ethical banking and corporate social responsibility (CSR) emerged. In the literature, in fact, there are three general approaches for banks to move towards sustainability. In the first approach, corporate social responsibility is considered as a path to sustainable banking. There are many articles in the literature encouraging banks to invest in CSR as Belasri et al. [44] claim that CSR has a positive impact on bank efficiency. In this regard, the results of the research of Nizam et al. [45], Maqbool and Zameer [46], Wu et al. [47], and Cornett, Erhemjamts, and Tehranian [48] explain that there is a positive relationship between the CSR activities and financial performance of banks. Nonetheless, Oyewumi et al. [49] argue that investing in CSR depletes the banks' financial resources, therefore, the recommendation that the banks to disclose their commitment to CSR activities to increase the banks' reputation, which will improve banks' financial performance.

In the second approach, banks are encouraged to carry out activities that directly reduce the harmful effects on the environment, mitigate carbon emissions, and protect the climate. Energy and water consumption reduction and waste management, for example, have been suggested for the service industry, especially banking, to be able to reduce their costs and present a good image of their brand in addition to fulfilling their environmental responsibility [50].

The third approach and trend in the literature for achieving sustainability goals for banks is to offer products that will contribute to sustainable development. Such research recommends that banks have value propositions that will lead to climate impact mitigation, adoption to the climate changes, reduction in environmental impact, eliminating hunger and poverty, increasing literacy, combating gender discrimination, etc.

Nwagwu [51] elaborates on the sustainable activities of the banks and their contribution to the Sustainable Development Goals (SDGs). He argues that the banks' policies can meet 6 out of 16 SDGs constituting no poverty, zero hunger, good health and wellbeing, quality education, gender equality, and clean water and infrastructure. Nwagwu [51] asserts that affordable services micro-loans and mobile banking lead to less poverty, accessible financial services to the farmers increase the yields and food security, which subsequently decreases hunger. The services that the banks provide such as savings opportunities and insurance empower the resilience of the households in dealing with the unexpected health problems which contribute to good health and wellbeing SDG. The banks' financial supports in education increases the quality of education. The banks' services empowering women's microcredit and financial literacy contribute to gender equality SDG. Nwagwu [51] debates that banks can contribute to clean water and infrastructure of SDG by providing the micro-loans and the other innovative products to public authorities to cope with the safe water. A close look at the Nwagwu [51] solutions reveals that the service and products a bank provides are the main sources the banks rely on to deal with the sustainability issues.

Nizam et al. [45] study 713 banks from 75 countries from 2013 to 2015. Their finding reveals that accessing environmental financing improve positively the financial performance of the banks. They argue that loan growth and quality management resulted from the environmental financing increase financial performance.

To provide a better understanding of the transition of banking toward sustainability, Seyfang, and Gilbert-Squires [52] develop a conceptual framework. According to this framework, the intersections between banking regimes and banking practices' transitions, firstly, should be identified and then constraints and obstacles to sustainable banking should be recognized to facilitate the transition toward sustainability.

Yip and Bocken [12] introduce sustainable business model archetypes for sustainability. They firstly identify eight sustainable business model archetypes using the literature and interviews with the experts which are: 1) maximizing material and energy efficiency, 2) substituting with digital processes, 3) encourage sufficiency, 4) adopting a stewardship role, 5) inclusive value creation, 6) repurposing for society/environment, 7) resilience in loan granting, 8) sustainable financial products. In the next step, they validated and tested the customer receptiveness for the archetypes. According to their results, the customers welcome more "substitute with digital processes", "adopt a stewardship role" and "encourage sufficiency" archetypes. They elaborate that substitute with digital processes archetype refers to the utilization of digital channels to reduce environmental impacts, the archetype of adopt a stewardship role outlines the importance of interacting actively with all stakeholders for awareness of their health and well-being, and the archetype of encourage sufficiency points out the solutions for reducing the need to the use of banking services.

Rebai, Azaiez, and Saidane [53] develop an index so-called the banking sustainability performance index (BSPI) to evaluate the sustainability performance in the banking sector. According to this index, the wellbeing of multiple stakeholders is targeted as the objective of the model. This index just considers the regulators and the civil society as the main stakeholders and ignores to evaluate the environmental effects of the banking activities. In this model, regulators by determining prudential norms strive to control the risk and ensure the stability of the financial system. Rebai et al. believe that the banks are the voice of civil societies and play a mediatory role in leading the other industries toward sustainable development. Since there is more than one stakeholder in this model, they applied the multi-attribute utility model to facilitate the performance evaluation based on a multi-objective context.

Raut, Cheikhrouhou, and Kharat [54] claim that corporate social responsibility alone is not the representation of sustainability performance. Therefore, they develop a multi-criteria decision-making model to assess sustainability performance in the banking sector. By incorporating the Balanced Scorecard, fuzzy Analytic Hierarchy Process (AHP), and fuzzy TOPSIS, Raut et al. [54] propose a four-dimension model for evaluation of the sustainability performance of banks. Whereas, financial stability, customer relationship management, internal business process, and environment-friendly management system constitute the four dimensions of this model.

On the other hand, the SBMs incorporate sustainability in the company's value proposition and value creation logic [55]. According to Nosratabadi et al. [1], sustainable business models constitute components that the interrelation among the components creates and delivers sustainable values for the multi-stakeholders. The SBMs generate competitive advantages to the firms and incorporate sustainable value proposition, sustainable value creation, and sustainable value capturing processes to bring economic benefits. In addition, Boons and Lüdeke-Freund [56] believe that in designing the financial model of an SBM, environmental and social impacts of the company must be taken into account. Despite the proven impact of the environmental performance on the performance among different industries and sectors, the role of sustainable business models on the environmental and financial performance of the banks is poorly understood. Therefore, the current study is conducted to develop a conceptual model for designing a sustainable business model in the banking sector. Hence, the study has three main objectives: 1) to identify the components of a sustainable business model for the banking sector, 2) to determine the importance of each of business model components in the sustainability performance of the banks, and 3) to analyze and to compare European banks' sustainability performance. Accordingly, this study strives to answer the following research questions:
- What are the components of a sustainable banking business model?
- How effective are the components of the sustainable banking business model in the sustainability performance of the banks?
- How sustainable are the European banking business models?

## 3. Research Methodology

In the methodology section, details of the sample are explained followed by an explanation of the phases of the study. Firstly, the detail of sampling and the banks and countries studied is provided. The Delphi method is applied to recognize the sustainable components of a bank business model followed by the use of an empirical assessment tool, Analytic Hierarchy Process (AHP), in order to prioritize the identified components. Besides, Lai, Wong, and Cheung [57] believe that the Application of AHP would be more acceptable over the Delphi method and there are many examples in the literature that utilized both methods to reach their goals. Lai et al. [57] results were in agreement with the findings of other studies (e.g., [58–62]). Each method is discussed in detail in the following sections.

*3.1. Sample Selection*

United Nations categorizes the European countries in Eastern Europe, Northern Europe, Southern Europe, Western Europe based on the geographic location of the countries. In this article, two countries from each region are selected Poland and Hungary from Eastern Europe, Norway and The UK from Northern Europe, Spain and Italy from Southern Europe, France and Germany from Western Europe are selected. For the Delphi method 16 experts from 16 private banks (two banks from each country) formed the expert panel. In this study, only Public Limited Company (PLC) commercial banks are investigated, because such banks, in comparison with the other banks, cover wider clients and provide wider ranges of services. Since in the initial contact with banks only two banks from some of the countries agreed to participate in the study, two banks from each country are selected. Whilst to get the second phase objective, an AHP questionnaire was distributed among the employees of the same banks used in the first phase to confirm and validate the finding of the study. It is worth mentioning that the data collection process took placed during the summer and fall of 2019. Due to privacy reasons, a coding system admitted instead of using the banks name in which each bank name constitutes three letters that the first letter is the first of the country name, the second letter is the letter B which stands for the word bank and a number which is either 1 or 2 that indicates that two banks are selected from each country. Therefore, PB1 and PB2 are the polish banks, HB1 and HB2 are two Hungarian banks, NB1 and NB2 are the Norwegian banks, BB1 and BB2 are the British banks, SB1 and SB2 are the Spanish banks, IB1 and IB2 are the Italian banks, FB1 and FB2 are the French banks, and finally GB1 and GB2 are the German banks which were investigated in this study. Figure 1 depicts the geographical location of the banks that are investigated in the current study. It is worth mentioning that only the banks located in the capitals of each of the mentioned countries were studied.

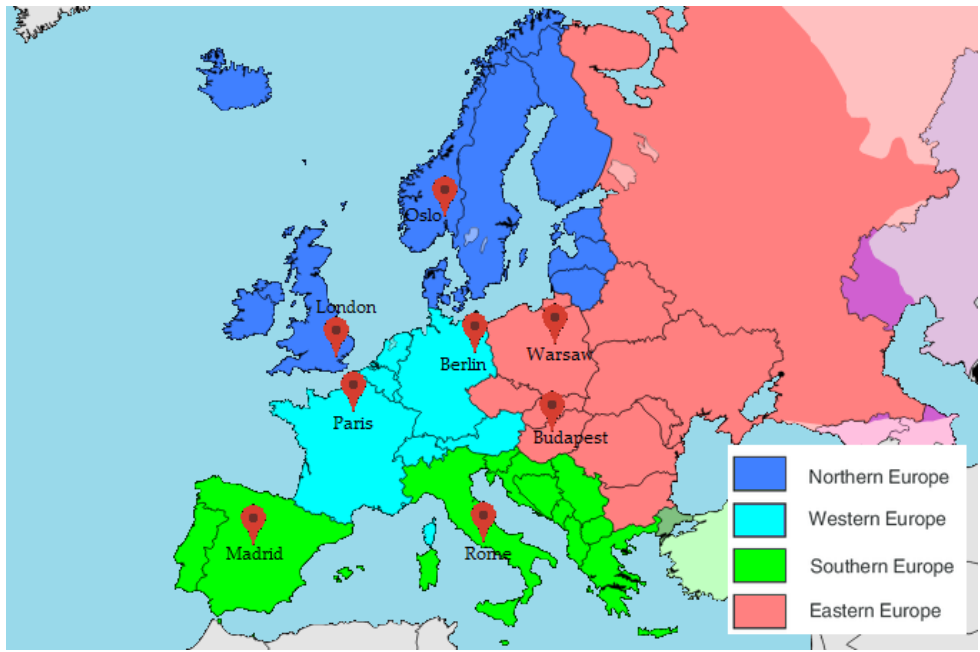

**Figure 1.** The geographic location of the sample of the study.

*3.2. Delphi Method*

The Delphi method is a brainstorming technique comprising iterative processes of collecting data from a panel of experts to find solutions opportunities [63]. The Delphi method, a structured communication technique, has three key characteristics. First, a panel comprised of knowledgeable experts who are purposely selected to give their opinion on a problem or situation. The second feature of this method is that all the panel members should remain unknown to each other in the entire data collection process. Third, in order to get feedback and develop a consensus among experts, group communication is necessary [64].

The RAND Corporation, in the 1950s, conducted studies to develop an approach in which the consensus of a group of experts is acquired. The Delphi method was the result of their work. Since then, the researchers utilize the method as an approach in different situations in which the consensus of experts is required [65]. Okoli and Pawlowski [66] argue that the Delphi method is applied for two main purposes: 1) forecasting and issue identification/prioritization and 2) developing a concept or a framework.

The Delphi method has widely used in the literature of business and management too. Adini et al. [67], for instance, provided a comprehensive framework enabling the organizations to assess and prioritize the concepts, approaches, and practices for resilience management. Tseng and Bui [68] apply the Delphi method to identify the main eco-innovation features for increasing industrial symbiosis performance. Barnes and Mattsson [69] apply a four-stage Delphi study with 25 experts to identify the main stimulators, barriers, and also future developments in collaborative consumption over the next ten years. Nambisan et al. [70] administered the Delphi method to provide a taxonomy of organizational design actions to increase the propensity of technology users to innovate in information technology. Czinkota and Ronkainen [71] utilize this method to predict the changes in the international business environment and their effects on corporate practices over the next decade. Of course, the mentioned studies are not the only works that have applied this method in their research. They are only examples of this research. It is worth mentioning that we use a panel of 16 experts from 16 banks investigated in this study to share their opinion on the sustainable elements of a business model in context to the banking sector.

*3.3. Analytic Hierarchy Process (AHP)*

Multi-criteria decision making (MCDM) methods deal with the selection of the best alternative among several potential options based on several criteria or attributes [72,73]. Therefore, the current study employed the Analytic Hierarchy Process (AHP), which is one of the MCDM, to prioritize the importance of SBBM components. In fact, this method gives priority to the SBBM components based on their effect on sustainability. The AHP has remarkably contributed to the literature of sustainability as, for instance, it is applied in common areas of sustainability and food, agriculture, urban development, fuel/biofuel, Energy [74]. Vaidya and Kumar [75] provide a comprehensive literature review of the applications of the AHP method. They categorized the research articles applied AHP into ten main categories: 1) selection, 2) evaluation, 3) cost-benefit, 4) allocation, 5) planning and development, 6) priority and ranking, 7) decision making, 8) forecasting, 9) medicine, and 10) quality function deployment.

In this method, the options are evaluated based on criteria to satisfy a goal. As is depicted in Figure 2, the AHP method includes three-level of goals, criteria, and alternatives where, ultimately, the methodology ranks the decision alternatives [76]. Hence, the first step is to define the problem according to the AHP model (Figure 2). Forming the paired comparison matrix is the next step of executing the AHP. In this step, the experts are asked to rank and compare the alternative in pairs according to their importance in achieving the goal. The experts can give a rank between 1 to 9 to show their priority where 9 represents extremely preferred, 7 refers to very strongly preferred, 5 stands for strongly preferred, 3 and 1 respectively point out to moderately preferred and equally preferred. While, 2, 4, 6, and 8 express the intermediate values of the adjacent judgments.

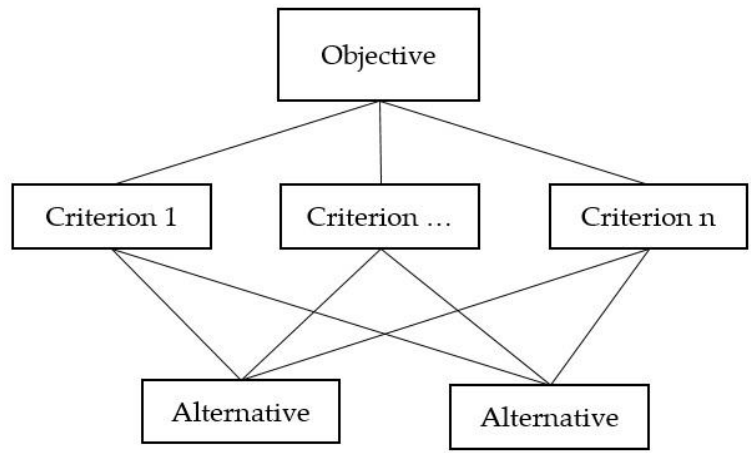

**Figure 2.** Analytic Hierarchy Process.

The next step in the AHP method is to design the pairwise comparison matrix in which the alternatives are judged based on the criteria (see Equation 1).

$$D = \begin{bmatrix} X_{11} & \cdots & X_{1n} \\ \vdots & \ddots & \vdots \\ X_{n1} & \cdots & X_{nn} \end{bmatrix} \quad (1)$$

To evaluate the reliability of the AHP method, the Consistency Ratio (CR) is evaluated. The CR value can be between 0 to 1 where 0 stands for the perfectly consistent judgments indicating the highest reliability of the results. There is a rule of thumb in the literature that CR should be around 0.10 [77], however, in some studies, the CR value of 0.2 was considered acceptable [78,79]. In this study, the CR value of 0.12 is considered acceptable and those questionnaires did not meet this requirement were excluded from further analysis. Equation 2 illustrates how to measure CR.

$$CR = \frac{CI}{RI} \quad (2)$$

where CI is Consistency Index and RI is the Random Index selecting using the Saaty scale [80]. CI can be obtained through Equation 3. To measure CI, it is needed to measure $\lambda\ max$ at first. The Equation 4 shows how to measure $\lambda\ max$.

$$CI = \left(\frac{\lambda(\max) - n}{n - 1}\right) \qquad (3)$$

$$\lambda\ max = \sum_{i=1}^{n} \frac{di}{n} \qquad \text{Where } i = 1,2,\ldots n \qquad (4)$$

## 4. Results and Discussion

*4.1. Phase 1: Recognition of the Key Components of a Bank Business Model*

In order to recognize the sustainable components of an efficient bank business model, a thorough review of business models and banking literature was undertaken. Despite there are frameworks to design a business model in the current literature, there is a deficiency in existing business model literature in context to the banking industry. Therefore, this study will, in some way, go towards filling that gap and also aims to provide a "bigger picture" for future research. The landmark components of the business models proposed by various authors are presented in Table 1. Finding reveals that there are 22 studies that tried to present frameworks to evaluate and design a business model. These models constitute 123 components in total, in which 24 components were narrowed down by eliminating the duplicate ones. These 24 business model components are summarized in Table 1. The third column of Table 1 refers to the authors who were using that component in their own proposed model.

**Table 1.** The business model elements in the literature and their source.

| Row | Components | Sources |
|---|---|---|
| 1 | Value proposition | [81–97] |
| 2 | Financial domain | [81,82,84–86,88,89,92–94,98–102] |
| 3 | Business processes | [82,83,87–89,92,93,95–97,100,101] |
| 4 | Distribution channel | [82,83,86–88,91,93,97,98,100,101] |
| 5 | Market segment | [84,85,92,94,97,99,100] |
| 6 | Core competencies | [81,84,85,87,96,99] |
| 7 | Supply chain management | [83,87,98,101,102] |
| 8 | Resources | [83,84,91,92,98] |
| 9 | Value chain structure | [84,92,94,102] |
| 10 | Customer interface | [85,90,96] |
| 11 | Strategy | [85,94,101] |
| 12 | Partner Network | [87,93,97] |
| 13 | Organizational form | [97,98] |
| 14 | Governance form | [95,99] |
| 15 | Market communication | [91,93] |
| 16 | Technology | [89] |
| 17 | Competitive position | [100] |
| 18 | Empowered employee | [87] |
| 19 | Mission | [92]. |
| 20 | Value exchange | [97] |
| 21 | Market model | [86] |
| 22 | Implementation model | [86] |
| 23 | Thread model | [86] |
| 24 | Knowledge management | [83] |

The Delphi method was used to determine the most important business model components using a panel of 16 experts, consisting of a senior manager from each of the 16 banks. Each member was aware that they would complete a task in conjunction with other bank managers, but they were not aware of the identity of the other panel members. Each member of the panel was individually given literature relating to business models and definitions for the 24 identified components. Each individual identified the components they considered important to the banking sector. The opinions of each were passed onto the other panel members via the researcher who facilitated the process. They, in turn, refined their lists taking into consideration the views of the others. In total, five iterations of reviews took place. Nine factors were identified as being the most important: 1) value proposition, 2) financial aspects, 3) business processes, 4) core competencies, 5) resources, 6) customer interface, 7) partner networks, 8) technology, and 9) target customers. A schematic of the Delphi processes applied in this study is shown in Figure 3.

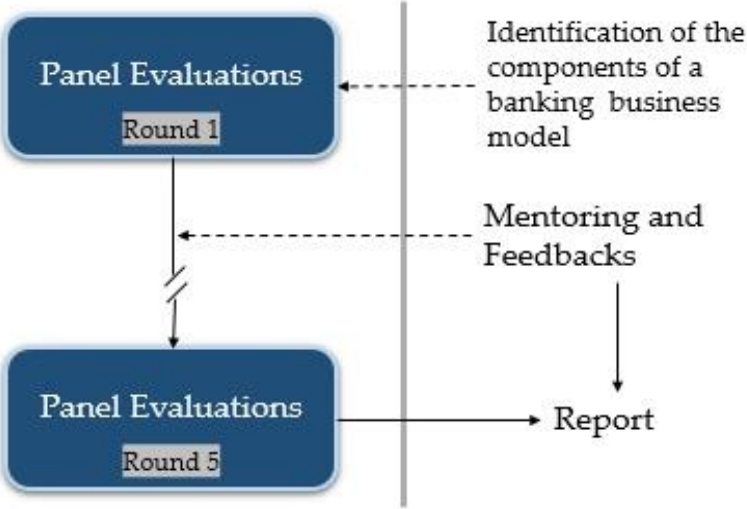

**Figure 3.** The processes of the Delphi method for identification of the components of the banking business model.

Where "value proposition" is described as a comprehensive assortment of a company's products and services [88], whereas "financial component" comprises two indices of the cost structure, which sums up the monetary consequences of the means employed in the business model and revenue model which describes the way a company makes money through a variety of revenue flows [88], "business processes" explains all the interconnected entities which create value in a business model [82], "core competencies" outlines the competencies necessary to execute the company's business objectives [88], "resources" describes sources which are used in value creation and value offering [83,84,91,92,98], "customer interface" is about the ways the business can reach customers [82], "partner networks" element explains the network of cooperative agreements with other companies necessary to efficiently offer and commercialize value [88], "technology" describes the technical functionality required to realize the service [89], and ultimately "target customers" refers to the segments of customers a company wants to offer value to [88]. These nine components make up the business model for the banking sector and are depicted in Figure 4. This concludes the first phase of the study and addresses the first research question. In phase two, already addressed components are assessed and prioritized according to the experts' opinions.

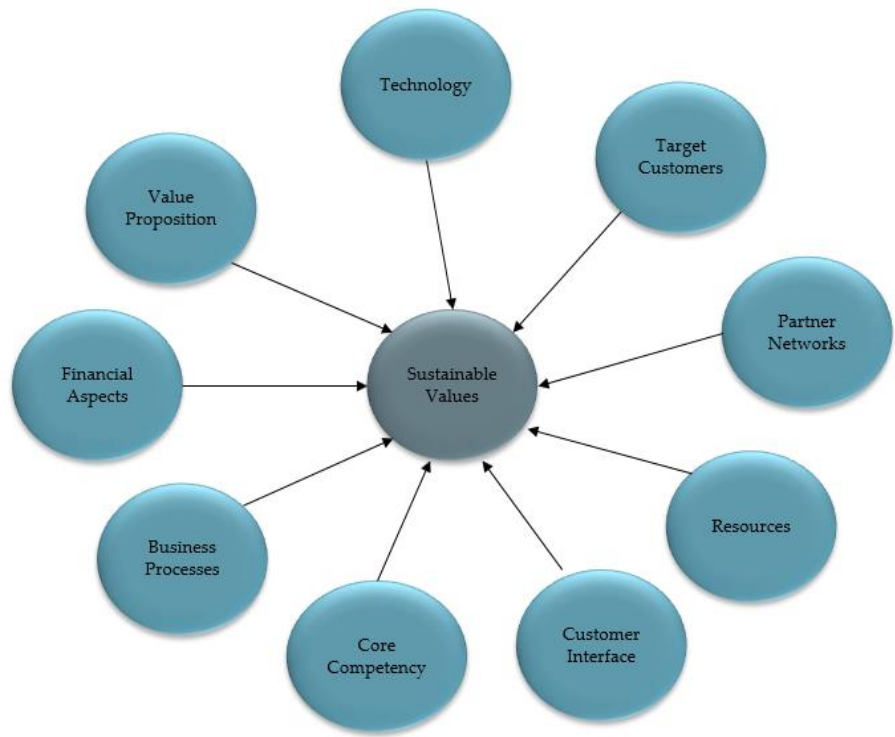

**Figure 4.** The proposed framework of the study for the Banking sustainable business model.

*4.2. Phase 2: Prioritization of the Bank Business Model Components*

The AHP method was utilized to prioritize the bank business model's components based on their importance in sustainability. To do so, an AHP questionnaire was designed based on the nine business model components. This questionnaire was distributed among 166 managers and employees of all 16 banks in Europe. According to Table 2, 166 questionnaires distributed among employees of 16 banks, and 147 useable responses were completed giving an acceptable response rate of 88%. Table 2 shows the number of respondents from each bank.

**Table 2.** The countries and banks surveyed in this study.

| Region | Country | Bank | No of questionnaires distributed | Usable completed questionnaires | Response rate |
|---|---|---|---|---|---|
| Northern Europe | Norway | NB1 | 9 | 8 | 87% |
| | | NB2 | 11 | 10 | 92% |
| | The UK | BB1 | 9 | 8 | 89% |
| | | BB2 | 8 | 7 | 89% |
| Eastern Europe | Poland | PB1 | 9 | 8 | 89% |
| | | PB2 | 12 | 11 | 89% |
| | Hungary | HB1 | 9 | 8 | 89% |
| | | HB2 | 13 | 11 | 84% |
| Western Europe | France | FB1 | 10 | 9 | 89% |
| | | FB2 | 10 | 9 | 90% |
| | Germany | GB1 | 12 | 11 | 90% |
| | | GB2 | 10 | 9 | 88% |
| Southern Europe | Spain | SB1 | 9 | 8 | 84% |
| | | SB2 | 13 | 12 | 89% |
| | Italy | IB1 | 10 | 9 | 89% |
| | | IB2 | 12 | 11 | 90% |
| Total | | 16 | 166 | 147 | 88% |

The employees were asked to rate the importance of each component against another component (first v's second) in terms of its importance in sustainability, with 9 being the most important scale. For example, Table 3 illustrates two lines from the questionnaire which showcase that value proposition is rated higher as compared to financial aspects for sustainability, while business processes are placed slightly above than value proposition.

Table 3. Example of questionnaire responses.

| First component | Preferred rates Paired comparisons based on sustainability | | | | | | | | | | | | | | | | | | Second component |
|---|---|---|---|---|---|---|---|---|---|---|---|---|---|---|---|---|---|---|---|
| Value Proposition | 9 | 8 | 7 | 6 | 5 | 4 | 3 | 2 | 1 | 2 | 3 | 4 | 5 | 6 | 7 | 8 | 9 | Financial Aspect |
| Value Proposition | 9 | 8 | 7 | 6 | 5 | 4 | 3 | 2 | 1 | 2 | 3 | 4 | 5 | 6 | 7 | 8 | 9 | Business Process |

Data collected from all 147 responses were analyzed using Expert Choice 11.0 software that performs pairwise comparisons based on the AHP methodology. The first round of analysis consisted of 147 rounds, one for each completed questionnaire. For every individual response weight associated with the components and the CR are calculated. Of the 147 completed questionnaires, 130 met the CR requirement of 0.12 and were included for further analysis. In the next step, the geometric mean of all the weighs related to each component were calculated. These values were entered into Expert Choice for the final round of analysis. The output is the final weight for each business model component. Each weigh indicates the importance of that component in achieving the objectives of sustainability. The weightings are given in Table 4.

Table 4. Priority ranking and weight of bank business model components.

| Priority Ranking | Business Model Components | Weight |
|---|---|---|
| 1 | Value proposition | 0.129 |
| 2 | Core competencies | 0.127 |
| 3 | Financial aspects | 0.123 |
| 4 | Business processes | 0.120 |
| 5 | Target customers | 0.113 |
| 6 | Resources | 0.110 |
| 7 | Technology | 0.109 |
| 8 | Customer interface | 0.094 |
| 9 | Partner networks | 0.075 |
| Total | | 1.000 |

The role of a business model is to explain how an entity creates value and generates profits to sustain itself in a competitive market. Similarly, the bank managers were requested to rate these nine business model components based on their importance in sustainability. The result illustrates that the bank managers rank these nine component of business model as follow: 1) value proposition (weighting 0.129), 2) core competencies (0.127), 3) financial aspects (0.123), 4) business processes (0.120), 5) target customers (0.113), 6) resources (0.110), 7) technology (0.109), 8) customer interface (0.094) and 9) partner network (0.075).

In other words, the component weights indicate that value proposition (0.129), core competencies (0.127), financial aspects (0.123) are the most important pillars of a bank's business model in terms of sustainability while partner networks (0.075) was rated as the least important element among the other 9 elements in the banking industry.

According to Wu and Shen [41], financial motivations are the main stimulations for the banks to adopt sustainability principles. A close look at the common literature of sustainability and banking discloses that sustainability is perceived from the financial stakeholders' and society's perspective and provided solutions are toward reducing and control the financial risk for the stakeholders and meeting the society benefits [103]. For instance, Nwagwu [51], Seyfang and Gilbert-Squires [52], Rebai et al. [53], and Raut et al. [54] propose frameworks to evaluate sustainability performance the banks whilst they just consider the clients' and the other stakeholders' benefits. Nizam et al. [45] and Yip and Bocken [12] are the only researchers have considered the environmental impact of the banking activities in their model where, for example, Yip and Bocken [12] recommend a so-called 'substitute with digital processes' business model archetype to utilize digital channels to reduce environmental impacts. In contrast, the proposed framework of the current study provides a tool for the banks' managers to design their business models in such a way that they can satisfy environmental, social and economic stakeholders' benefits at the same time. Since the concept of the business model covers four main aspects of a business naming value proposition, value creation, value delivering and financial models, designing a sustainable business model provide the opportunity to the managers to implement the sustainability principles in each of the processes of value-creating, value delivering, value proposition and value capturing (financial aspects) - not only in the financial aspects.

All business entities have their competitive business model, although not every enterprise knows what exactly the distinctive features it possesses [104]. Therefore, this prototype can help bank managers to determine their needs and adopt a model which, in the long run, meets the pre-defined objectives of the company. They can keep themselves up-to-date and be innovative by analyzing the business model components that help in value appreciation and financial stability. While the term 'business model' was once an abstract idea, it is now a more concrete tool that can be used to evaluate business performance. This model can be used as a benchmark to assess a bank's performance, internally as well as externally, in a timely manner.

*4.3. Phase 3: Comparing the Performance of Banks Across Europe*

In this section, the sixteen banks studied in this research are compared and evaluated based on their performance in each component of the business model. To do so, the same expert panels were requested to calculate their bank performance on each of the business model elements.

The results of the comparison of the banks with respect to all the business model components and the priority calculated for each of these banks relative to the above elements are presented in Table 5. Table 5 shows that GB1 outperformed other banks in terms of a value proposition. In other words, the performance of the German bank 1 in achieving the sustainability goals of the value proposition has been better than the other.

**Table 5.** The Sustainability of the Business Model Components of the Banks Studied.

|  | Value Proposition | Core Competency | Financial Aspects | Business Process | Target Customers | Resources | Technology | Customer Interface | Partner Network | Business model |
|---|---|---|---|---|---|---|---|---|---|---|
| NB1 | 0.067 | 0.063 | 0.068 | 0.064 | 0.066 | 0.063 | 0.064 | 0.069 | 0.067 | 0.066 |
| NB2 | 0.062 | 0.064 | 0.066 | 0.061 | 0.068 | 0.061 | 0.065 | 0.061 | 0.061 | 0.063 |
| BB1 | 0.064 | 0.065 | 0.066 | 0.061 | 0.057 | 0.063 | 0.069 | 0.059 | 0.059 | 0.063 |
| BB2 | 0.054 | 0.062 | 0.064 | 0.068 | 0.058 | 0.059 | 0.067 | 0.059 | 0.059 | 0.061 |
| PB1 | 0.063 | 0.059 | 0.059 | 0.058 | 0.059 | 0.059 | 0.061 | 0.066 | 0.056 | 0.060 |
| PB2 | 0.057 | 0.066 | 0.059 | 0.064 | 0.063 | 0.063 | 0.066 | 0.062 | 0.069 | 0.063 |
| HB1 | 0.069 | 0.058 | 0.067 | 0.060 | 0.069 | 0.063 | 0.059 | 0.065 | 0.059 | 0.063 |
| HB2 | 0.068 | 0.061 | 0.062 | 0.064 | 0.065 | 0.059 | 0.059 | 0.062 | 0.059 | 0.062 |
| FB1 | 0.054 | 0.060 | 0.058 | 0.063 | 0.058 | 0.060 | 0.066 | 0.061 | 0.063 | 0.060 |
| FB2 | 0.067 | 0.056 | 0.066 | 0.065 | 0.066 | 0.068 | 0.058 | 0.059 | 0.066 | 0.063 |
| GB1 | 0.070 | 0.062 | 0.059 | 0.061 | 0.063 | 0.068 | 0.062 | 0.066 | 0.065 | 0.064 |
| GB2 | 0.066 | 0.069 | 0.064 | 0.062 | 0.059 | 0.061 | 0.063 | 0.059 | 0.065 | 0.063 |
| SB1 | 0.067 | 0.062 | 0.062 | 0.062 | 0.062 | 0.060 | 0.062 | 0.066 | 0.059 | 0.062 |
| SB2 | 0.054 | 0.068 | 0.061 | 0.062 | 0.062 | 0.069 | 0.065 | 0.067 | 0.064 | 0.064 |
| IB1 | 0.055 | 0.060 | 0.053 | 0.063 | 0.065 | 0.061 | 0.056 | 0.056 | 0.067 | 0.060 |
| IB2 | 0.063 | 0.065 | 0.066 | 0.062 | 0.060 | 0.063 | 0.058 | 0.063 | 0.062 | 0.062 |
| Total | 1.000 | 1.000 | 1.000 | 1.000 | 1.000 | 1.000 | 1.000 | 1.000 | 1.000 | 1.000 |

According to Table 5, the performance of German bank 2 in core competencies in achieving the sustainability goals was higher than the other banks in other countries. The results revealed that NB1, the Norwegian bank 1, best performed in the financial aspect. In other words, NB1 presented a better performance in executing strategies for sustainable revenue generation compared to the other studied banks. BB2, which is the UK bank 1, reported a higher performance in implementing sustainable business processes. HB1, which stands for Hungary's bank 1, performed better in the target customer than other banks. Among the banks studied in southern Europe, SB2 is the only bank to perform better in a business model component than other banks in other European regions. SB2 performed better in the provision of sustainable resources to run its business model. Likewise, BB1, NB1, and PB2 had the highest performance in achieving sustainability goals in the components of technology, customer interface, and partner network, respectively.

The last column of Table 5 represents the overall performance of each bank's business model in achieving sustainability goals. The business model of Norwegian bank 1 acquired a higher score in being sustainable. Figure 5 presents an opportunity to compare the sustainable performance of different banks' business models. In terms of sustainability, the business model of the Polish bank1, the French bank 1, and the Italian bank 1 showed the lowest sustainable performance among the other banks studied.

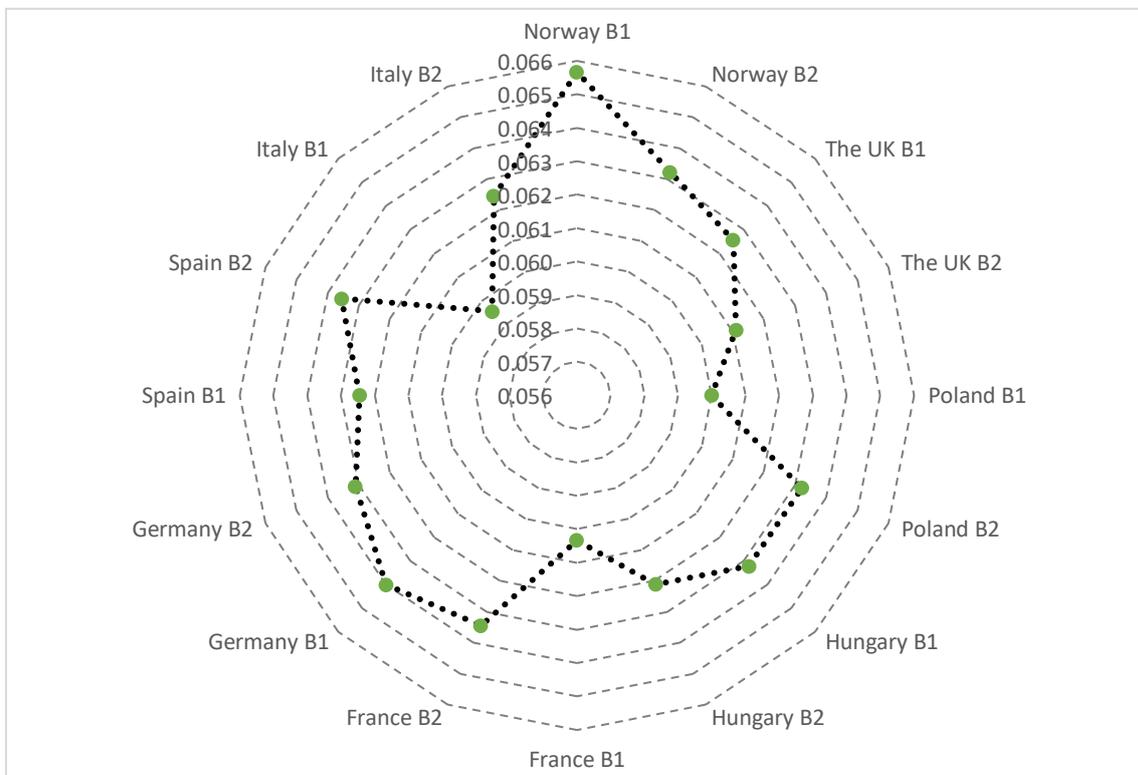

**Figure 5.** Comparison of the sustainability performance of the business model of the banks studied.

The arithmetic mean for the sustainability performance of the business model of the two banks in each country was measured, and the results are shown in Figure 6. The arithmetic means value for the sustainability of the Norwegian and German banks' business models with a value of 0.064 is higher than the rest of the countries. This indicates that the banks studied in these two countries have a more sustainable business model compared to the banks of other studied countries. Likewise, the studied banks of Hungary and Spain, with an arithmetic mean of 0.063, came in second in the sustainability of the business model. The banks of the UK, Poland, and France with an arithmetic mean of 0.062, ranked third in the sustainability of the business model. Finally, the two Italian banks scored the lowest on their business model sustainability compared to other banks.

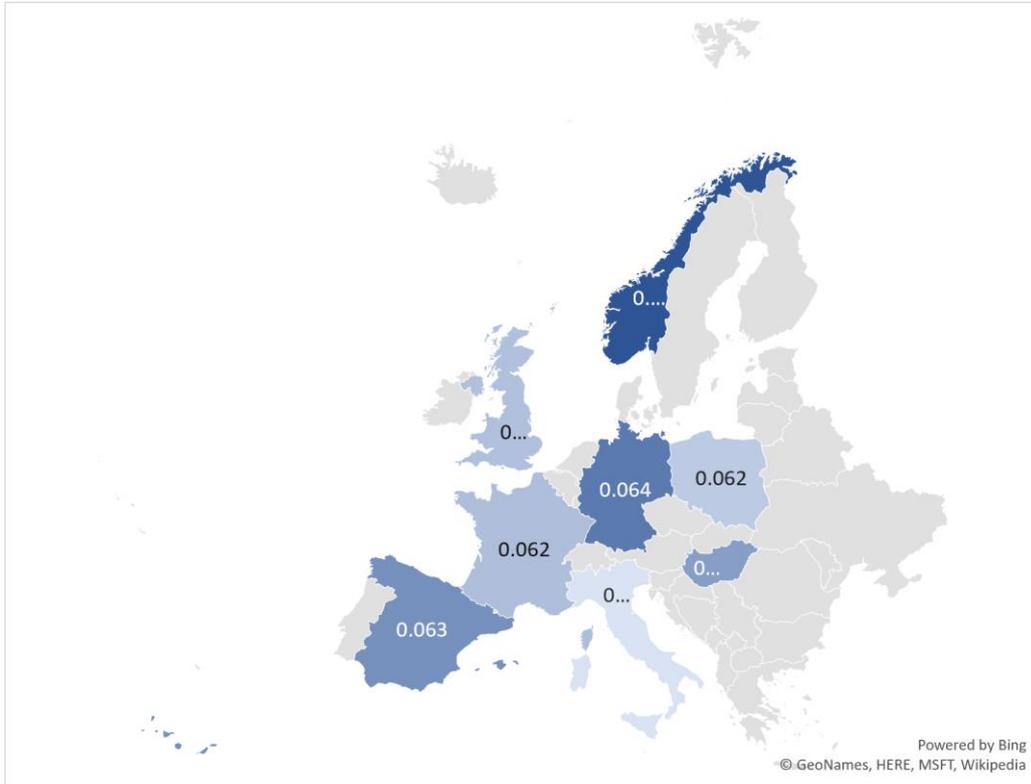

**Figure 6.** The sustainability of the business model of the banks of the countries studied in this article.

## 5. Conclusion

Admitting being responsible for the environment and society while making a profit, stimulating the business to be more innovative in resource management and value creation. Creating a value that is able to unleash opportunities to meet the environmental benefit and to provide money is very challenging for the businesses. Since the European countries, which are leading to sustainability, have a very sophisticated financial system, the evaluation of the sustainability level of their banks is essential. Therefore, the main objective of the current study was to assess the sustainability performance of banks in Europe. The finding of the current study provides guidelines for the mangers in the banking industry to design a sustainable business model in which all the trends and the behavior of every single component can be monitored and adjusted regarding their effect on the sustainability performance of the banks. Previously, the financial aspect of a business model was the only criteria to evaluate the effectiveness and efficiency of the banks' business model. These models failed to provide a comprehensive overview of value creation and value delivery processes.

The authors feel that this current study has successfully able to bridge the gap in the literature in context to the banking industry. This was achieved by adopting Delphi and AHP methodology and participation of a panel of experts from sixteen banks from eight European countries including Norway, the UK, Hungary, Poland, Germany, France, Spain, and Italy. The experts were first asked to identify sustainable business model components and then prioritized those components according to their professional experience. The study resulted in a niche business model for the banking sector, which incorporates core components of sustainability. These components are value proposition (weighting 0.129), 2) core competencies (0.127), 3) financial aspects (0.123), 4) business processes (0.120), 5) target customers (0.113), 6) resources (0.110), 7) technology (0.109), 8) customer interface (0.094), and 9) partner network (0.075). An in-depth appraisal of this model helps the users to evaluate options as needed to ensure that the bank remains competitive, sustainable, and profitable. The results of the comparison of the banks studied by each country showed that the sustainability of the Norwegian and German banks' business models is higher than the rest of the countries. The studied banks of Hungary and Spain came in second in the sustainability of the business model. The banks of the UK, Poland, and France ranked third in the sustainability of the business model. Finally, the two Italian banks scored the lowest on their business model sustainability compared to other banks.

Although the banking industry is a matured industry and banking systems are almost similar all over the world, a generalization of the finding is hard as the data collected from only two banks from each country. Therefore, it is recommended that future researchers consider similar research to study all banks in one country to give a better picture of the sustainability performance of business models in that country. On the other hand, this study provides a theoretical ground for future studies to develop sustainable business models for banking sectors. Hence, it is recommended for future studies to design a sustainable banking business model on the basis of the proposed model of this study.

**Author Contributions:** Conceptualization, S.N., A.M.; methodology, S.N., A.M.; software, S.N., A.M., S.S.; validation, S.N., A.M., G.P., S.S.; formal analysis, S.N., A.M., G.P., S.S.; investigation, S.N., A.M., G.P., S.S.; data curation, S.N., A.M., G.P., S.S.; writing—original draft preparation, S.N., A.M., G.P., S.S.; writing—review and editing, S.N., A.M., G.P., S.S.; visualization, S.N., A.M., G.P., S.S.; supervision, S.S.; controlling and confirming the results, S.S; project administration, A.M.

**Funding:** We acknowledge the financial support of this work by the Hungarian State and the European Union under the EFOP-3.6.1-16-2016-00010 project and the 2017-1.3.1-VKE-2017-00025 project.

**Conflicts of Interest:** The authors declare no conflict of interest.